\documentstyle[graphicx]{europhys}

\def\And{{\rm and\ }}

\def\stars{\bigskip\centerline{***}\medskip}

\newif\ifboo \boofalse

\def\Review#1{\boofalse{\it #1},}
\def\Name#1{{\sc #1},}
\def\Vol#1{\ifboo Vol. {\bf #1}\else{\bf #1}\fi}
\def\Year#1{\ifboo #1\else(#1)\fi}

\def\Page#1{\ifboo {\rm p. #1}\else{\rm #1}\fi}

\begin{document}
\euro{}{}{}{}
\Date{}
\shorttitle{Charge fluctuation effects in NaV$_{2}$O$_{5}$}
\title{Electron-phonon and spin-phonon coupling in
       NaV$_{2}$O$_{5}$: charge fluctuations effects}
\author{E.Ya. Sherman\footnote{Permanent address: Moscow
        Institute of Physics and Technology, 141700, Dolgoprudny, Moscow Region, Russia},
       M. Fischer, P. Lemmens, P. H. M. van Loosdrecht, and G.
       G\"untherodt}
\institute{II. Physikalisches Institut, RWTH-Aachen, 52056 Aachen,
           Germany}
\rec{}{}
\pacs{
   \Pacs{63}{20Kr}{Phonon-electron and phonon-phonon interactions}
   \Pacs{75}{30Et}{Exchange and superexchange interactions}
   \Pacs{78}{30$-$j}{Infrared and Raman spectra}
}
\maketitle
\begin{abstract}
We show that the asymmetric crystal environment of the V site in
the ladder compound NaV$_{2}$O$_{5}$ leads to a strong coupling of
vanadium 3$d$\ electrons to phonons. This coupling causes
fluctuations of the charge on the V ions, and favors a transition
to a charge-ordered state at low temperatures. In the low
temperature phase the charge fluctuations modulate the spin-spin
superexchange interaction, resulting in a strong spin-phonon
coupling.
\end{abstract}
\section{Introduction}
The spin-Peierls transition in one-dimensional spin systems is one
of the best known and investigated cooperative phenomena in
low-dimensional physics. This transition comprises a
dimerization of the lattice of a Heisenberg spin chain,
leading to an alternation of the exchange parameter $J$.
As a result, nearest-neighbor spins form a singlet state
accompanied by the opening of a spin gap.
Consequently, the static spin susceptibility drops drastically below the
transition temperature ($T_{\mathrm{SP}}$).
This effect, predicted by Pytte \cite{Pytte}, was first observed in one-dimensional
organic compounds \cite{Bray}. The discovery was followed by theoretical work of
Cross and Fisher \cite{Cross}, who showed a possibility to calculate
$T_{\mathrm{SP}}$ from the coupling of spins to the lattice.
The discovery  of the spin-Peierls transition in the inorganic
spin-chain compound CuGeO$_{3}$ \cite{Hase} renewed interest in this phenomenon.
In this half-filled chain compound each Cu site holds
one ''hole'' with spin $s=1/2$. The spin-spin exchange interaction
occurs via oxygen orbitals.
Lattice vibrations cause a modulation of the spin-spin superexchange
due to changes in the inter-ionic distances and bond angles. This leads to
a spin-lattice coupling which plays an important role as
a driving force of the spin-Peierls transition.

The ladder system NaV$_{2}$O$_{5}$ was proposed to be a second
inorganic spin-Peierls compound \cite{Tc}. It contains VO$_5$
pyramids which form V-O ladders extended along the $b-$axis \cite{Capry}, see
Fig.~1a. For half the ladders the apical oxygen O3 is
above the V ions while for the other half they are below them.
From the chemical formula one finds an average vanadium valence of
+4.5. This implies that a 3$d$-electron is shared between two V
ions within one rung, making NaV$_{2}$O$_{5}$ a ''quarter-filled''
ladder compound \cite{Smolinski}. The shared electron introduces an
additional degree of freedom: there is a possibility for the
electron to move between the right ($R$) and left ($L$) V ions
within one rung. This feature leads to physical differences
between NaV$_{2}$O$_{5}$ and CuGeO$_{3}$, the most striking one
being a possible $charge$ ordering in NaV$_{2}$O$_{5}$. If an
additional degree of freedom is involved in a phase transition,
one should expect enhancement of the anomalies accompanying it. In
fact, available experimental data on NaV$_{2}$O$_{5}$ reveal
strong anomalies in heat conductivity \cite{Vasiliev}, specific
heat \cite{Hemberger}, inelastic neutron scattering
\cite{neutron}, and inelastic light scattering \cite{Fischer}.
These observations can be considered as (i) confirmation of
changes in the electronic subsystem in addition to a spin gap
opening, and (ii) indirect evidence for charge ordering. Direct
evidence of charge separation within a rung was provided by a
nuclear magnetic resonance experiment \cite{NMR} which showed that
two non-equivalent V ions exist below $T_{C}$. In addition, X-ray
and elastic neutron scattering revealed a lattice superstructure
with a modulation vector $(1/2,1/2,1/4)$ in the low-temperature
phase \cite{Xray}. However, the nature of the charge ordering and
the strength of the charge asymmetry are not clear yet.

\begin{figure}[htb]
\centering{\includegraphics[width=0.95\textwidth,clip=true]{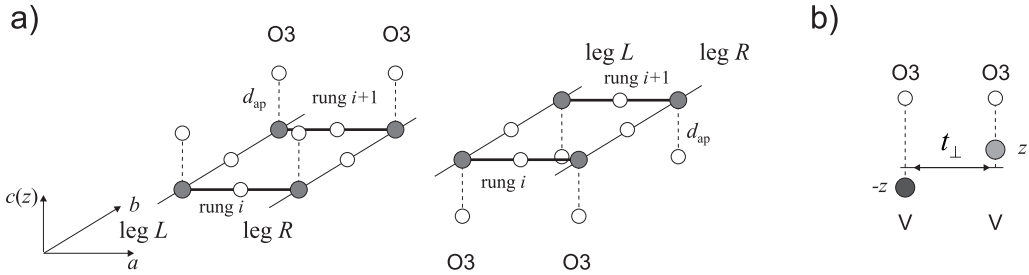}}
\caption{a) Schematic plot of a part the crystal
structure of the NaV$_{2}$O$_{5}$ compound. Grey circles denote V
ions while open circles correspond to the O ions. The V-O-V
rungs are shown in bold. b) Charge disproportion arising due to
out-of-phase V ion displacements along the $z$-axis. Light-grey
(dark-grey) circle corresponds to decreased (increased) electronic
density on the site.} \label{fig1}
\end{figure}

Two different models for the charge ordered phase were
introduced recently, notably an ''in-line'' \cite{Fulde} and
a ''zigzag'' structure \cite{Seno,Mostovoy}.
In the in-line structure all V ions within one leg have
the same charge $Q_{il}$ but $Q_{iL}\neq Q_{iR},$ where $i$ enumerates
rungs along the ladder, while $l=L,R$ denotes the sites within one rung.
For the zigzag state one has $Q_{iR}=Q_{i+1,L}$.
For both the states the electron moves within one rung, {\it i.e.} $Q_{iL}+Q_{iR}=1$.
It was proposed in \cite{Fulde} that the phase transition occurs
in two stages. First the charge subsystem orders, which is followed
by the opening of a spin gap. Mostovoy and Khomskii \cite{Mostovoy}
argued that a zigzag ordering itself may lead to the opening of a
spin gap due to an induced alternation of the exchange
within the ladders.
Nevertheless, coupling  to the lattice is always important for
the spin-Peierls transition either as a driving force or as a factor
stabilizing the charge ordering \cite{Mostovoy}.
For a better understanding of the role the lattice plays it is important to
know the values and physical origin of the parameters involved in a
model taking the spin, charge and lattice degrees of freedom into account.

Here we investigate a mechanism of electron-phonon coupling
in NaV$_{2}$O$_{5}$ and determine the scale of the parameters
of the spin-phonon interaction. Due to the asymmetric structural environment
of the V sites (see Fig.~1a) the vibration-induced shifts of
electron site energies are proportional to $z$-axis
displacements of the ions. The small V-O3 equilibrium distance
$d_{\mathrm{ap}}\approx 1.61\ \mathrm{\AA}$ \cite{Smolinski}
leads to a strong contribution of the O3 ion
Coulomb field, $E_{\mathrm{ap}}$, to the 3$d$ electron site energy.
A phonon mode involving a non-equivalent
modulation of $d_{\mathrm{ap}}$ on the right and left side
therefore causes a charge redistribution between the V ions on a
rung. It will be shown that in the charge-ordered
phase this effect leads to a strong spin-phonon coupling
activated by the charge disproportion. The Coulomb field enhances the
role of electronic correlations proposed to be responsible for the charge
ordering \cite{Fulde,Seno, Mostovoy} and leads to phonon-driven fluctuations
of the exchange. Finally, two possible experimental
consequences of the charge disproportion and fluctuations
for optical spectroscopy will be analyzed.
\section{One-rung charge disproportion}
Consider a V-O-V rung, denoting the indirect hopping between
the V ions as $t_{\perp }\approx0.35 $ eV \cite{Smolinski,Horsch}.
Then there are two in-rung states separated in energy by $2t_{\perp }$:
an odd and an even one, and both V ions have charge $+4.5$.
Now consider an out-of-phase vibration of the two V ions along the $z$-axis
(see Fig.~1b). The corresponding zero-point amplitude is
$z_{0}=\sqrt{\hbar /M\Omega }$, where $M$ is the V ion mass,
and $\Omega $ is the phonon frequency. For $\Omega =400$ cm$^{-1},$ which
is an estimate justified later by our experimental data,
$z_{0}\approx $ 0.04 \AA. Under a displacement ($\pm z$)
of the V ions  the on-site energies will be
changed by an amount $\pm Cz$, inducing a charge transfer between
the V ions ($C$ is the deformation potential).
By determining the eigenstates of this asymmetric two-level system
the ground state energy $\varepsilon _{0}$ and amount of the
charge transfer is easily calculated, yielding
\begin{equation}
\varepsilon _0=-\sqrt{t_\perp^2+C^2z^2},\qquad
\delta Q=Q-\frac{1}{2}=\pm\frac{Cz}{2\sqrt{t_\perp^2+C^2z^2}},
\end{equation}
where $Q=0$ for V$^{+5}$ state.
A point-charge Coulomb estimate of $E_{\mathrm{ap}}=2e/d_{\mathrm{ap}}^{2}$
yields $C=eE_{\mathrm{ap}}=12$ eV/\AA.
Therefore $Cz_0/t_\perp$, the parameter characterizing the
electron-phonon coupling,  is of order unity.
Assuming that an adiabatic approximation is valid
($\Omega\ll t_\perp$), and calculating the mean value with the
vibrational ground state wave function
$\left| \phi \right\rangle =(\pi z_0^2)^{-1/4}\ e^{-z^2/2z_0^2}$,
we obtain the
charge fluctuations in the charge-symmetrical (high-temperature) phase:
\begin{equation}
\left\langle\phi\right|\left(Q-1/2\right)^2\left|\phi\right\rangle
= \frac{1}{4}-\frac{\sqrt{\pi}\xi}{4z_0}e^{\xi^2/z_0^2}
\left[1-\Phi(\xi/z_0)\right] ,
\end{equation}
where $\xi\equiv t_\perp/C=0.03$ \AA,
and $\Phi(\xi/z_0)$ is the error function.
Since $\xi \approx z_0$, we obtain
$\left\langle \phi \right|(Q-1/2)^2\left| \phi\right\rangle \approx 0.18$,
{\it i.e.} the fluctuations are of the
order of the net average charge.

Now we can estimate the role of the displacement-induced site asymmetry in
the formation of the low-temperature phase.
Suppose that inter-site Coulomb interactions, leading to intersite
correlations, induce a static splitting $2\Delta _{\mathrm{cor}}$ of the
on-site levels within a rung. This splitting causes a charge
disproportion $\varphi=|Q_{L}-Q_{R}|$.
This purely correlation-induced disproportion $\varphi _{\mathrm{cor}}$
is of the order of $\Delta _{\mathrm{cor}}/t_{\bot }$.
However, the apical oxygen field at the $d-$electron
considerably enhances this charge redistribution. The correlation-induced
charge asymmetry will cause a difference in the Coulomb forces acting on
the V ions. This difference induces static opposite displacements of the ions,
leading to a further increase of the charge disproportion. Eventually, this process
will stop due to elastic energy of the lattice. The additional splitting 2$\Delta _{\mathrm{lat}}$
of the on-site levels leads to a total site energy shift
$\pm\Delta=\Delta _{\mathrm{lat}}+\Delta _{\mathrm{cor}}$,
the average ion charges are given by
$Q_{L,R}-1/2=\pm \Delta /2\left(t_{\perp }^{2}+\Delta ^{2}\right) ^{1/2}$.
The new equilibrium positions $\overline{z}$ of the V ions
are determined by a self-consistency equation:
\begin{equation}
x=\frac{C^2}{2t_\perp\widetilde{k}}\cdot \frac{1+x}{\sqrt{1+\tau^2(1+x)^2}}\ ,
\end{equation}
where $\tau\equiv\Delta _{{\rm cor}}/t_{\perp }$,
$\Delta _{{\rm lat}}\equiv C\overline{z}$,
$x\equiv \Delta _{{\rm lat}}/$ $\Delta _{{\rm cor}}$,
and $\widetilde{k}$ the non-$d-$electron contribution to the
total force constant $k=M\Omega ^{2}$ ($\approx 30\textrm{ eV/\AA}^2$)
relevant for the vibration.
Eq.~(3) is a consequence of the equilibrium condition
$\widetilde{k}\overline{z}=C\varphi/2$ and Eq.~(1) for the charge
transfer between the V ions. It has the following limiting cases for $\tau \ll 1$:
\begin{equation}
x=\left\{
\begin{array}{lll}
A/\left(1-A\right),    &  & 1-A^2\gg\tau ^2 \\
                       &  &                 \\
2^{1/3}\tau ^{-2/3}-1, &  & 1-A^2\ll\tau ^2
\end{array}
\right.\ ,
\end{equation}
with $A\equiv C^{2}/2t_{\perp }\widetilde{k}$. The first limit
corresponds to a response of the lattice to a relatively small
correlation-induced charge disproportion. The second one shows
that if the lattice is close to an instability
($A\approx 1$) even weak correlations induce a large charge
disproportion.  The $d$-electron contribution to the force
constant can be calculated from Eq.~(1) as
$k_{d}=\partial^{2}\varepsilon_{0}/2\partial z^{2}\approx -C^{2}/2t_{\perp }$
$(\approx -200\textrm{ eV/\AA}^2$ for $C=eE_{{\rm ap}})$,
yielding $A=-k_{d}/(k-k_{d})\approx 0.85$.
It follows from Eq.~(4) that $x$ can be larger than
unity, in which case the main contribution to $\varphi$ comes from
the apical oxygen (O3) field.
This field enhances the correlation induced asymmetry of the V ions
and thus favors charge ordering.

For $\tau >1$ Eq.~(3) yields $x\sim 1/\tau$. The charge disproportion
and ion displacement now
saturate at  $\varphi\approx 1$ and
$\overline{z}\sim C/\widetilde{k}$, respectively.
The corresponding $\overline{z}$ is of the order of $z_{0}$, in
agreement with the experimental shifts reported in Ref. \cite{Smaalen}.
The corresponding site energy shift is of the order of
$\overline{z}C\sim t_{\perp }$. Within
a linear approximation the phonon-induced charge transfer in the ordered
phase can be written as:
\begin{equation}
\delta Q=C\delta z\frac{t_{\perp }^{2}}{\left( t_{\perp }^{2}+\Delta
^{2}\right) ^{3/2}},
\end{equation}
where $\delta z$ is the out-of-phase displacement from the new
equilibrium positions. As we will see below, these charge fluctuations lead
to a spin-phonon coupling in the low-temperature phase.
\section{Phonon-induced fluctuations of the exchange}
It has been proposed that above $T_{c}$ the spin system of
NaV$_{2}$O$_{5}$ can be mapped onto an antiferromagnetic Heisenberg
chain with $\widehat{H}=J\sum {\mathbf s}_{i}{\mathbf s}_{i+1}$,
where ${\mathbf s}_{i}$ is the spin {\em per rung} \cite{Smolinski,Horsch}
and $J\approx 560$ K \cite{Tc}.
Using this idea we can calculate $J$ within a simple model. The
superexchange occurs mainly due to hopping via the oxygen orbitals within
the legs. We denote the hopping integral between V and O orbitals along the
ladder as $t_{pd}$, and the one-electron energy difference
for the in-leg transition O(2$p$)$\rightarrow$V(3$d$) transition
as $\varepsilon\approx 6.5$ eV \cite{Horsch}. Allowing for
inequivalent nearest-neighbor rungs $i$ and $i+1$, $J$ can be expressed
as
\begin{equation}
J=2\sum_{l=L,R}\frac{t_{i,l}^{2}t_{i+1,l}^{2}}{\varepsilon ^{2}t_{\perp }},
\end{equation}
where the $t_{i,l}$ and $t_{i+1,l}$ are hoppings of a hole from the V
sites to the O ion between them. The values of $t^2_{i,l}$ being
proportional to the probability to find the hole at a
corresponding V site can be readily expressed via the charges of
the ions: $t_{i,l}^{2}=t_{pd}^{2}Q_{i,l}$. From Eq.~(6) we obtain
$J=2t_{pd}^{4}(Q_{i,L}Q_{i+1,L}+Q_{i,R}Q_{i+1,R})/\varepsilon ^{2}t_{\perp }$%
\footnote{Expressing of $J$ as a function of the charges is a
common feature of the approaches based on perturbation theory
\cite{Gros}.}. The charge-ordering modified exchange
$\widetilde{J}$ for the in-line and zigzag phases is uniform and
may be written as
\begin{equation}
\widetilde{J}=\frac{J}{t_{\perp }^{2}+\Delta ^{2}}\cdot \left\{
\begin{array}{lll}
t_{\perp }^{2}, &  & \mbox{zigzag phase},Q_{i,L}=Q_{i+1,R} \\
&  &  \\
2\Delta ^{2}+t_{\perp }^{2}, &  & \mbox{in-line phase},Q_{i,L}=Q_{i+1,L}
\end{array}
\right. .
\end{equation}
The charge separation in the zigzag phase decreases the exchange while the
in-line ordering increases it \cite{Gros}. These effects can be easily
understood since in the zigzag phase the electrons are more separated in space
while in the in-line phase they are more close to each other.
The experimentally found $\widetilde{J}/J\approx 0.8$ \cite{Weiden}
thus is indicative of a zigzag type of  charge ordering in NaV$_2$O$_5$.

The phonon causing charge fluctuations also modulates the hopping
$t_{i,l}$ and, in turn, the exchange $\widetilde{J}$.
For phonon momentum $q$ along the ladder the modulation
of $\widetilde{J}$ is found to be
\begin{equation}
\frac{\delta \widetilde{J}}{\delta z}=4J\left( Q_{L}-Q_{R}\right) f(q)\frac{%
\delta Q}{\delta z},
\end{equation}
with $f(q)=i\sin (qd/2)$ and $f(q)=\cos (qd/2)$ for the zigzag and
in-line phases, respectively. $\delta Q$ is the charge transfer in
Eq.~(5) and $d$ the in-ladder lattice constant. This contribution
to the spin-phonon coupling is solely due to charge asymmetry and
therefore a characteristic feature of the low-temperature phase.
Following the approach of Pytte \cite{Pytte} we estimate the
spin-phonon coupling constant as $\lambda \sim z_{0}^{2}(\delta
\widetilde{J}/\delta z) ^{2}(\Omega \widetilde{J})^{-1}$. Note
that, within the present model, a complete charge ordering
($\varphi=1$,$\Delta\gg t_\perp$) suppresses spin-phonon coupling
since it prohibits phonon-induced charge fluctuations (see
Eq.~(5)). From Eqs.~(5)-(8) we obtain:
\begin{equation}
\lambda \sim \frac{C^{2}z_{0}^{2}}{t_{\perp }^{2}}\frac{J}{\Omega }\frac{%
\Delta ^{2}}{\left( \Delta ^{2}+t_{\perp }^{2}\right) ^{3}}.
\end{equation}
The maximal $\lambda $ is reached at $\Delta =t_{\perp }/\sqrt{2}$,
corresponding to $\varphi =1/\sqrt{3}$, $\widetilde{J}/J=2/3$, and $\lambda
\sim 1$.

Finally we address the influence of charge ordering on possible
frustrating next-nearest-neighbor interactions ($J^\prime$).
This is interesting since frustration ($J^\prime/J$) may be a driving force
for the opening of a spin gap.
If we suppose that both nearest and next-nearest couplings
are determined by the charge order, we obtain for the zigzag phase:
$\widetilde{J}=J(1-\varphi ^{2})$ \cite{Gros} and
$\widetilde{J}^{\prime }=J^{\prime }\left[ (1-\varphi)^{2}+(1+\varphi )^{2}\right]
/2=J^{\prime }(1+\varphi ^{2})$. Zigzag ordering apparently leads to an
enhancement  of the role of frustration.
\section{Experimental consequences}
The effects of the phonon-induced charge fluctuations and
charge disproportion directly influence the electronic and lattice properties of the
crystal and should therefore manifest themselves in Raman and infrared spectroscopy.

One experimental consequence of the charge disproportion
considered above is that the out-of-phase vibration of the V ions
should become Raman active below $T_c$. The normal vectors of even
(in-phase) and odd vibrations of the V ions (see Fig.~1b) can be
written as ${\mathbf u}_{\mathrm{g}}=(1,1)/\sqrt{2}$ and ${\mathbf
u}_{\mathrm{u}}=(1,-1)/\sqrt{2}$, respectively. If $\varphi \neq 0$,
these vibrations are mixed and their eigenvectors can be
written in the form:
\begin{equation}
\mathbf{u}_{\mathrm 1}=\mathbf{u}_{\mathrm{g}}\cos \theta +\mathbf{u}_{\mathrm{u}}\sin \theta ,
\qquad
\mathbf{u}_{\mathrm 2}=-\mathbf{u}_{\mathrm{g}}\sin \theta + \mathbf{u}_{\mathrm{u}}\cos \theta.
\end{equation}
The angle $\theta$ is determined by the charge disproportion
within a rung ($\theta =0$ at $T>T_{c}$). There are two reasons
for Raman activity of the second vibration in Eq.~(10). The first
one is the presence of the Raman active ${\bf u}_{\rm g}$
component in the ${\bf u}_{\rm 2}$ mode. The second reason is that
charge asymmetry leads to electronically different V ions
(V$^{+4.5+\varphi/2}$ and V$^{+4.5-\varphi/2}$). Therefore,
displacements of these ions give different contributions to the
Raman polarizability, leading to Raman activity of the
${\bf u}_{\mathrm{u}}$ mode. These two effects, which can not be
separated presently, lead to an intensity $I_2$ of the new mode
proportional to $\sim\varphi^2$.

Experimental Raman data for $T=100$~K and $T=5$~K are presented in
Fig.~2. At $T=100$~K, only a single Lorentzian mode at 420 cm$^{-1}$
is observed. Below the phase transition, at $T=5$~K, this mode
splits in two components centered at $394$~cm$^{-1}$ and 429~cm$^{-1}$,
respectively. The lower energy peak appearing for $T<T_c$ is attributed
to the $\mathbf{u}_{\mathrm 2}$ vibration in Eq.~(10). The ratio of the
integrated intensities of the ''new''  (394 cm$^{-1}$) and ''old''
(429 cm$^{-1}$) phonons is $I_2/I_1\approx0.4$. This is
considerably larger than for the phonons arising solely due the
superstructure formation, where the ratio is found to be smaller
than 0.1 \cite{Fischer1}. In principle the ratio $I_2/I_1$ could
be used to estimate the extend of the charge ordering ($\varphi\sim\sqrt{I_2/I_1}$).
However, the reliability of this estimate would be questionable
since the Raman polarizability of the $\mathbf{u}_{\mathrm{u}}$ is
not only determined by charge ordering but is also influenced by resonance enhancement
effects\cite{konst}.

\begin{figure}[htb]
\centering{\includegraphics[width=0.50\textwidth,clip=true]{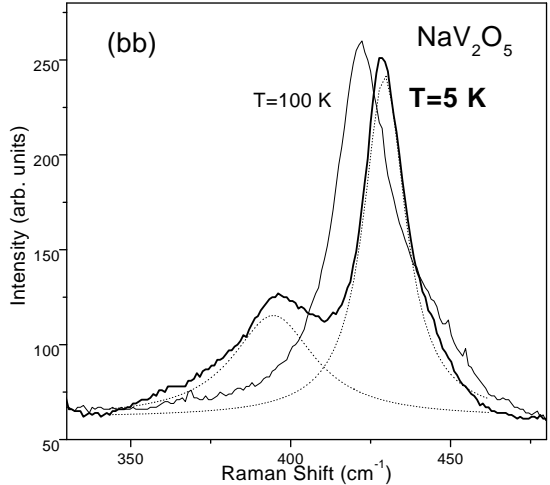}}
%
\caption{Raman spectra of NaV$_2$O$_5$ at $T=100$~K (thin line)
and $T=5$~K (bold line) demonstrating the
appearance of a new intensive phonon at 394 cm$^{-1}$ in the
low-temperature phase. Polarizations of the incident and scattered
light are parallel to the ladders. The dashed line shows
Lorentzian fits to the $T=5$~K spectrum.} \label{fig3}
\end{figure}

Another experimental consequence is related to charge fluctuations
as observed in infrared (IR) experiments.
IR absorption spectra at $T=300$ K show a peak at
$\hbar\omega _{\max }\approx $1~eV \cite{IR}. This peak has been
assigned to in-rung electronic transitions.
The in-rung hopping was estimated there as $t_{\perp }\approx 0.3$ eV,
whereas the difference $\hbar\omega _{\max }-$ 2$t_{\perp }\approx 0.4$ eV
was interpreted as a result of an in-rung charge separation already
at $T>T_{c}$. Here, we propose that this shift may be due to the
electron-phonon coupling considered above, even {\em without}
static charge separation.
In the presence of the out-of-phase phonon mode the splitting of the
one-rung levels is $2\sqrt{t_{\perp }^{2}+C^{2}z^{2}}>2t_{\perp }$.
Taking into account that the probability of a $z-$axis
displacement is determined by the vibrational wave function
$\left| \phi \right\rangle$, we obtain a band in the absorption spectra
which starts at $2t_{\perp }$ and peaks at
$\hbar\omega _{\max }\approx 2t_{\perp}+\left\langle \phi \right| C^{2}z^{2}/t_{\perp }\left| \phi \right\rangle$.
Taking the same set of parameters as before, we get
$\hbar\omega _{\max }-$ 2$t_{\perp}\approx C^{2}z_{0}^{2}/t_{\perp }\approx 0.6$ eV,
in good agreement with the experiment.
Based on this model we expect an increase of $\hbar\omega _{\max }$
in the low-temperature phase, as well as a narrowing of the peak due to the
suppression of charge fluctuations by the charge ordering.
\section{Conclusion}
We investigated electron- and spin-phonon coupling in
NaV$_{2}$O$_{5}$ caused by the structural vanadium site asymmetry.
The electron-phonon coupling leads to charge fluctuations, and considerably
increases a correlation induced charge disproportion.
The fluctuations induce a \textit{strong} spin-phonon
coupling in the charge-ordered phase. We stress, however,
that this interaction is \textit{not} the driving force of the transition.
The charge ordering at low temperatures is manifested in Raman scattering by a
new intense phonon peak. The phonon-induced charge fluctuations in the
high-temperature phase may explain the absorption peak observed
near 1 eV in IR experiments.

\stars

Valuable discussions with C.~Pinettes and  M.~Grove are gratefully
acknowledged. This work is supported by the DFG through SFB341,
INTAS grant 96-410, and BMBF project 13N6586/8. E.Ya.S. acknowledges
support from the AvH Foundation.

\begin{thebibliography}{99}
\bibitem{Pytte}
  \Name{E. Pytte}
  \Review{Phys. Rev. B} \Vol{10} \Year{1974} \Page{4637}.
\bibitem{Bray}
  \Name{J.W. Bray, H.R. Hart Jr., L.V. Interrante, I.S. Jacobs,
        J.S. Kasper, G.D. Watkins, S.H. Wee \And J.C. Bonner}
  \Review{Phys. Rev. Lett.} \Vol{35} \Year{1975} \Page{744}.
\bibitem{Cross}
  \Name{M.C. Cross \And D.S. Fisher}
  \Review{Phys. Rev. B} \Vol{19} \Year{1978} \Page{402}.
\bibitem{Hase}
  \Name{M. Hase, I. Terasaki, \And K. Uchinokura}
  \Review{Phys. Rev. Lett.} \Vol{70} \Year{1993} \Page{3651}.
\bibitem{Tc}
  \Name{M. Isobe \And Y. Ueda}
  \Review{J. Phys. Soc. Jpn.} \Vol{65} \Year{1996} \Page{1178}.
\bibitem{Capry}
  \Name{P. A. Capry \And J. Galy}
  \Review{Acta Crist. B} \Vol{31} \Year{1975} \Page{1481}
\bibitem{Smolinski}
  \Name{H. Smolinski, C. Gros, W. Weber, U. Peuchert, G. Roth,
        M. Weiden \And  C. Geibel}
  \Review{Phys. Rev. Lett.} \Vol{80} \Year{1998} \Page{5164}.
\bibitem{Vasiliev}
  \Name{A.N. Vasil'ev, V.V. Pryadun, D. Khomskii, G. Dhalenne,
        A. Revcolevschi, M. Isobe \And Y. Ueda}
  \Review{Phys. Rev. Lett.} \Vol{81} \Year{1998} \Page{1949}.
\bibitem{Hemberger}
  \Name{J. Hemberger, M. Lohmann, M. Nicklas, A. Loidl, M. Klemm, G. Obermeier
        \And S. Horn}
  \Review{Europhys. Lett.} \Vol{42} \Year{1998} \Page{661}.
\bibitem{neutron}
  \Name{Y. Fujii, H. Nakao, T. Yosihama, M. Nishi, K. Nakajima,
        K. Kakurai, M. Isobe, Y. Ueda \And H. Sawa}
  \Review{J. Phys. Soc. Jpn.} \Vol{66} \Year{1997} \Page{326}.
\bibitem{Fischer}
  \Name{M. Fischer, P. Lemmens, G. G\"{u}ntherodt, A. Mischenko,
        M. Weiden, G. Geibel \And F. Steglich}
  \Review{Phys. Rev. B} \Vol{58} \Year{1998} \Page{14159}.
\bibitem{NMR}
  \Name{T. Ohama, H. Yasuoka, M. Isobe \And Y. Ueda}
  \Review{Phys. Rev. B} \Vol{59} \Year{1999} \Page{3299}.
\bibitem{Xray}
  \Name{T. Yosihama, M. Nishi, K. Nakajima, K. Kakurai, M. Isobe,
        C. Kagami \And Y. Ueda}
  \Review{J. Phys. Soc. Jpn.} \Vol{67} \Year{1998}  \Page{744}.
\bibitem{Fulde}
  \Name{P. Thalmeier \And P. Fulde}
  \Review{Europhys. Lett.} \Vol{44} \Year{1998} \Page{242}.
\bibitem{Seno}
  \Name{H. Seo \And H. Fukuyama}
  \Review{J. Phys. Soc. Jpn.} \Vol{67} \Year{1998} \Page{2602}.
\bibitem{Mostovoy}
  \Name{M. Mostovoy \And D.I. Khomskii} cond-mat/9806215 Preprint \Year{1998}.
\bibitem{Horsch}
  \Name{P. Horsch \And F. Mack}
  \Review{Europ. Phys. J. B} \Vol{5} \Year{1998} \Page{367}.
\bibitem{Smaalen}
  \Name{J. L\"{u}decke, A. Jobst, S. van Smaalen, E. Morre,
        C. Geibel \And H.-G. Krane}
  \Review{Phys. Rev. Lett.} \Vol{82} \Year{1999} \Page{3633}.
\bibitem{Gros}
  \Name{C. Gros and R. Valent\'\i}
  \Review{Phys. Rev. Lett.} \Vol{82} \Year{1999} \Page{976}.
\bibitem{Weiden}
  \Name{M. Weiden, R. Hauptmann, C. Geibel, F. Steglich, M. Fischer,
        P. Lemmens \And G. G\"{u}ntherodt}
  \Review{Z. f\"{u}r Physik B} \Vol{103} \Year{1997} \Page{1}.
\bibitem{Fischer1}
  \Name{M. Fischer, P. Lemmens, E. Sherman, G. G\"{u}ntherodt,
        M. Weiden, G. Geibel \And F. Steglich}
  \Review{Phys. Rev. B} \Vol{60} \Year{1999} \Page{7284}.
\bibitem{konst}
  \Name{M. J. Konstantinov\'c, K. Ladavac, A. Beli\'c,  Z. V. Popovi\'c,
        A. N. Vasil'ev, M. Isobe \And Y. Ueda}
  \Review{J. Phys. Cond. Matter} \Vol{11} \Year{1999} \Page{2103}.
\bibitem{IR}
  \Name{A. Damascelli, D. van der Marel, M. Gr\"{u}ninger, C. Presura,
        T.T.M. Palstra, J. Jedoudez \And A. Revcolevschi}
  \Review{Phys. Rev. Lett.} \Vol{81} \Year{1998} \Page{918}.
\end{thebibliography}
\end{document}